\documentstyle{article}

\newlength{\defaultparindent}
\newenvironment{Default Paragraph Font}{}{}
\input{tcilatex}

\begin{document}

$\backslash$$\backslash$

Title: Dimensionless Constants of the Fundamental Physical Interactions

viewed by the Model of Expansive Nondecelerative Universe

Authors: Jozef  Sima, Miroslav Sukenik and Magdalena
Sukenikova

Comments: 5 pages, AmsTex

Report-no: SS-99-03

$\backslash$$\backslash$

Abstract

A profound relationship between the ENU (Expansive Nondecelerative Universe)
and dimensionless constants of the fundamental physical interactions is
presented. The contribution corrects the Dirac presumption on a time
decrease of the gravitational constant $G$ and using simple relations it
precises the values of the vector bosons $x$ and $y$.

$\backslash$$\backslash$

\begin{center}
{\bf Dimensionless Constants of the Fundamental Physical Interactions}

{\bf viewed by the Model of Expansive Nondecelerative Universe}

Jozef Sima, Miroslav Sukenik and Magdalena Sukenikova

Slovak Technical University, Radlinskeho 9, 812 37 Bratislava, Slovakia
\end{center}

Abstract. A \ profound relationship between the ENU (Expansive
Nondecelerative Universe) and dimensionless constants of the fundamental
physical interactions is presented. The contribution corrects the Dirac
presumption on a time decrease of the gravitational constant $G$ and using
simple relations it precises the values of the vector bosons $x$ and $y$.

There is no problem in defining dimensionless physical constants, namely
that of strong interaction $\alpha _{s}$

$\alpha _{s}=1$ \ \ \ \ \ \ \ \ \ \ \ \ \ \ \ \ \ \ \ \ \ \ \ \ \ \ \ \ \ \
\ \ \ \ \ \ \ \ \ \ \ \ \ \ \ \ \ \ \ \ \ \ \ \ \ \ \ \ \ \ \ \ (1)

and electromagnetic interaction $\alpha _{e}$

$\alpha _{e}=\frac{e^{2}}{4\pi \epsilon _{o}.\hbar .c}=7.29\times 10^{-3}$ \
\ \ \ \ \ \ \ \ \ \ \ \ \ \ \ \ \ \ \ \ \ \ \ \ \ \ \ \ \ \ \ \ \ \ (2)

For the constant of weak interaction $\alpha _{w}$ various modes of
expression are used, the simplest of which is [1]

$\alpha _{w}=\frac{g_{_{F}}.m_{P}^{2}.c}{\hbar ^{3}}\approx 10^{-6}$ \ \ \ \
\ \ \ \ \ \ \ \ \ \ \ \ \ \ \ \ \ \ \ \ \ \ \ \ \ \ \ \ \ \ \ \ \ \ \ \ \ \
\ (3)

where $g_{F}$ is the Fermi constant ($g_{F}$ $\sim $ 10$^{-62}$ J.m$^{3}$)
and $m_{p}$ is the proton mass. Ambiguousness appears in the gravitational
constant $\alpha _{g}$ expression, the most familiar one being

$\alpha _{g}=\frac{G.m_{P}^{2}}{\hbar .c}\approx 10^{-39}$ \ \ \ \ \ \ \ \ \
\ \ \ \ \ \ \ \ \ \ \ \ \ \ \ \ \ \ \ \ \ \ \ \ \ \ \ \ \ \ \ \ \ \ (4)

The value of $\alpha _{g}$ depends on the mass option. As a matter of
tradition, the mass of proton $m_{P}$ is usually taken, there is, however,
no justified reason for it. On the other hand, there should be a certain
mass $m_{x}$, the introduction into $\alpha _{g}$ of which will be
rationalizable and justifiable. Gravitational forces are far-reaching,
theoretically boundless, forces. Due to the operation of the hierarchical
rotational gravitational systems (HRGS), the actual gravitational effects
are considered as being finite. Gravitation can manifest itself only where
density of the gravitational energy of HRGS is higher than the critical
density $\varepsilon _{crit}$. In the ENU model using the Vaidya metrics it
was shown [2] that

$\epsilon _{g}=-\frac{c^{4}.R}{8\pi .G}=-\frac{3m.c^{2}}{4\pi .a.r^{2}}$ \ \
\ \ \ \ \ \ \ \ \ \ \ \ \ \ \ \ \ \ \ \ \ \ \ \ \ \ \ \ \ \ \ \ \ \ \ \ \ \
\ \ (5)

where $\varepsilon _{g}$ is the density of gravitational energy created by a
body with the mass $m$ at the distance $r$, $R$ is the scalar curvature, $a$
is the gauge factor that at present

$a\approx 10^{26}m$ \ \ \ \ \ \ \ \ \ \ \ \ \ \ \ \ \ \ \ \ \ \ \ \ \ \ \ \
\ \ \ \ \ \ \ \ \ \ \ \ \ \ \ \ \ \ \ \ \ \ \ \ \ \ \ \ \ \ (6)

In the ENU model [2, 3] $\varepsilon _{crit}$ is expressed as

$\epsilon _{crit}=\frac{3c^{4}}{8\pi .G.a^{2}}$ \ \ \ \ \ \ \ \ \ \ \ \ \ \
\ \ \ \ \ \ \ \ \ \ \ \ \ \ \ \ \ \ \ \ \ \ \ \ \ \ \ \ \ \ \ \ \ \ \ \ \ \
\ \ (7)

In the cases when

$\left| \epsilon _{g}\right| =\epsilon _{crit}$ \ \ \ \ \ \ \ \ \ \ \ \ \ \
\ \ \ \ \ \ \ \ \ \ \ \ \ \ \ \ \ \ \ \ \ \ \ \ \ \ \ \ \ \ \ \ \ \ \ \ \ \
\ \ \ \ \ \ \ (8)

using equations (5) and (7) the following relation is obtained

$r=r_{ef(m)}=\left( a.R_{g(m)}\right) ^{1/2}$ \ \ \ \ \ \ \ \ \ \ \ \ \ \ \
\ \ \ \ \ \ \ \ \ \ \ \ \ \ \ \ \ \ \ \ \ \ \ \ \ \ (9)

in which $r_{ef(m)}$ represents the effective gravitational range of a body
with the mass $m$, $R_{g(m)}$ is its gravitational radius. Since for the
Compton wave $\lambda $\ of the particle with the mass $m$

$\lambda =\frac{\hbar }{m.c}$ \ \ \ \ \ \ \ \ \ \ \ \ \ \ \ \ \ \ \ \ \ \ \
\ \ \ \ \ \ \ \ \ \ \ \ \ \ \ \ \ \ \ \ \ \ \ \ \ \ \ \ \ \ \ \ \ \ \ \ \ \
\ (10)

a particle having the mass $m_{x}$ for which $\lambda $= $r_{ef(m)}$ must
exist. It thus represents the lighest particle able to gravitationally
influence its environment. Relating (9) and (10) leads to

$m_{x}=\left( \frac{\hbar ^{2}}{2G.a}\right) ^{1/3}$ \ \ \ \ \ \ \ \ \ \ \ \
\ \ \ \ \ \ \ \ \ \ \ \ \ \ \ \ \ \ \ \ \ \ \ \ \ \ \ \ \ \ \ \ \ \ \ \ \ \
(11)

At present

$m_{x}\approx 10^{-28}kg$ \ \ \ \ \ \ \ \ \ \ \ \ \ \ \ \ \ \ \ \ \ \ \ \ \
\ \ \ \ \ \ \ \ \ \ \ \ \ \ \ \ \ \ \ \ \ \ \ \ \ \ \ \ (12)

This value can be introduced into $\alpha _{g}$ without additional
presumptions and then

$\alpha _{g}=\frac{Gm_{x}^{2}}{\hbar .c}=\frac{m_{x}^{2}}{m_{Pc}^{2}}\approx
10^{-40}$ \ \ \ \ \ \ \ \ \ \ \ \ \ \ \ \ \ \ \ \ \ \ \ \ \ \ \ \ \ \ \ \ \
\ \ \ \ (13)

where $m_{Pc}$ is the Planck mass

$m_{Pc}\approx 10^{19}GeV$ \ \ \ \ \ \ \ \ \ \ \ \ \ \ \ \ \ \ \ \ \ \ \ \ \
\ \ \ \ \ \ \ \ \ \ \ \ \ \ \ \ \ \ \ \ \ \ \ \ \ \ (14)

In the ENU model [2,3] it was proved that

$a=c.t=\frac{2G.M_{u}}{c^{2}}$ \ \ \ \ \ \ \ \ \ \ \ \ \ \ \ \ \ \ \ \ \ \ \
\ \ \ \ \ \ \ \ \ \ \ \ \ \ \ \ \ \ \ \ \ \ \ \ \ \ \ (15)

where $t$ is the cosmological time, $M_{u}$\ is the mass of the Universe. It
follows from (11), (13) and (15)

$\alpha _{g}~(t)^{-2/3}$ \ \ \ \ \ \ \ \ \ \ \ \ \ \ \ \ \ \ \ \ \ \ \ \ \ \
\ \ \ \ \ \ \ \ \ \ \ \ \ \ \ \ \ \ \ \ \ \ \ \ \ \ \ \ \ \ \ \ \ (16)

It is worth pointing out at the importance of relation (16). It brings an
evidence that $\alpha _{g}$ is not a true constant but a time decreasing
quantity. This fact leads to some interesting consequences.

The Compton dimension of particle $m_{x}$ is about 10$^{-15}$ m and
corresponding $t_{x}$ $\sim $ 10$^{-23}$ s. Consequently, for the cosmologic
time $t$ expressed in $t_{x}$ units using (11), (13) and (15), relation (17)
can be obtained

$\frac{t}{t_{x}}\approx \frac{1}{\alpha _{g}}$ \ \ \ \ \ \ \ \ \ \ \ \ \ \ \
\ \ \ \ \ \ \ \ \ \ \ \ \ \ \ \ \ \ \ \ \ \ \ \ \ \ \ \ \ \ \ \ \ \ \ \ \ \
\ \ \ \ \ \ \ \ (17)

This fact was known to Dirac who, however, $t_{x}$ put equal to the atomic
time. It was, however, only a coincidence since at present the Compton
wavelength of the particle $m_{x}$ approaches the dimension of atomic
nucleus. Since Dirac introduced the constant mass of the proton when
expressing $\alpha _{g}$ and at the same time relied on the validity of
relation (17), he formulated a false conclusion on time decrease of the
gravitational constant $G$. In reality, $G$ is a true constant and $\alpha
_{g}$ is a time decreasing quantity (see relations 11, 13 a 16).

In the ENU model the Universe is mass-space-time (in the language of special
relativity) closed. In the framework of Friedmann model for such systems it
is supposed [1] that

$M_{u}\approx \frac{m_{P}}{\alpha _{g}^{2}}$ \ \ \ \ \ \ \ \ \ \ \ \ \ \ \ \
\ \ \ \ \ \ \ \ \ \ \ \ \ \ \ \ \ \ \ \ \ \ \ \ \ \ \ \ \ \ \ \ \ \ \ \ \ \
\ \ \ \ \ (18)

The above relation cannot be unambiguously proved, using (11), (13), (15) it
can be, however, derived that

$M_{u}\approx \frac{m_{x}}{\alpha _{g}^{2}}$ \ \ \ \ \ \ \ \ \ \ \ \ \ \ \ \
\ \ \ \ \ \ \ \ \ \ \ \ \ \ \ \ \ \ \ \ \ \ \ \ \ \ \ \ \ \ \ \ \ \ \ \ \ \
\ \ \ \ \ (19)

In addition, it holds in Friedmann model [1] that

$k.T\approx \left( \frac{\hbar ^{3}.c^{5}}{G.t^{2}}\right) ^{1/4}$ \ \ \ \ \
\ \ \ \ \ \ \ \ \ \ \ \ \ \ \ \ \ \ \ \ \ \ \ \ \ \ \ \ \ \ \ \ \ \ \ \ \ \
\ \ \ \ \ \ \ (20)

where $T$ is the temperature of the relict radiation. For closed systems it
is supposed [1] that

$k.T_{\min }\approx \alpha _{g}^{1/4}.m_{p}.c^{2}$ \ \ \ \ \ \ \ \ \ \ \ \ \
\ \ \ \ \ \ \ \ \ \ \ \ \ \ \ \ \ \ \ \ \ \ \ \ \ \ \ \ \ \ \ \ \ \ (21)

\bigskip Relation (21) cannot be derived, however, it follows from (11),
(13) and (20) that

$k.T\approx \alpha _{g}^{1/4}.m_{x}.c^{2}$ \ \ \ \ \ \ \ \ \ \ \ \ \ \ \ \ \
\ \ \ \ \ \ \ \ \ \ \ \ \ \ \ \ \ \ \ \ \ \ \ \ \ \ \ \ \ \ \ \ \ \ (22)

If in a known empirical formula [1]

$H\approx \frac{\alpha _{g}.m_{P}.c^{2}}{\hbar }$ \ \ \ \ \ \ \ \ \ \ \ \ \
\ \ \ \ \ \ \ \ \ \ \ \ \ \ \ \ \ \ \ \ \ \ \ \ \ \ \ \ \ \ \ \ \ \ \ \ \ \
\ \ \ \ (23)

where $H$ is the Hubble constant, the mass $m_{P}$ is substituted for $m_{x}$%
, relation (24) emerges which can be derived directly using (11), (13) and
(15)

$H\approx \frac{\alpha _{g}.m_{x}.c^{2}}{\hbar }$ \ \ \ \ \ \ \ \ \ \ \ \ \
\ \ \ \ \ \ \ \ \ \ \ \ \ \ \ \ \ \ \ \ \ \ \ \ \ \ \ \ \ \ \ \ \ \ \ \ \ \
\ \ \ \ (24)

It should be worth mentioning that relations (19), (22) and (24) are in fact
closely connected to dependence (17) which can be directly derived from
them. It is obvious that the introduction of $m_{x}$ into the ``constant'' $%
\alpha _{g}$ enables to formulate known empirical formulae as
self-sustaining, rationalizable and derivable relations that are or should
be valid in our Universe.

Due to the time decreasing of $\alpha _{g}$ (16) there had to be the time in
the past when

$\alpha _{g}=\alpha _{s}$ \ \ \ \ \ \ \ \ \ \ \ \ \ \ \ \ \ \ \ \ \ \ \ \ \
\ \ \ \ \ \ \ \ \ \ \ \ \ \ \ \ \ \ \ \ \ \ \ \ \ \ \ \ \ \ \ \ \ \ \ \ (25)

It follows of (1), (13) and (25) that at that time

$m_{Pc}=m_{x}\approx 10^{19}GeV$ \ \ \ \ \ \ \ \ \ \ \ \ \ \ \ \ \ \ \ \ \ \
\ \ \ \ \ \ \ \ \ \ \ \ \ \ \ \ \ \ \ \ \ (26)

It really happened at the beginning of the Universe expansion ($t$ $\sim $ 10%
$^{-43}$ s).

At about $t$ $\sim $ 10$^{-40}$ s following the beginning of the expansion
the quantities $\alpha _{e}$ and $\alpha _{g}$ were of identical value

$\alpha _{e}=\alpha _{g}$ \ \ \ \ \ \ \ \ \ \ \ \ \ \ \ \ \ \ \ \ \ \ \ \ \
\ \ \ \ \ \ \ \ \ \ \ \ \ \ \ \ \ \ \ \ \ \ \ \ \ \ \ \ \ \ \ \ \ \ \ \ (27)

and, consequently, treatment of (2), (13) and (27) leads to relation

$m_{x}=m_{Pc}.\alpha _{e}^{1/2}\approx 10^{18}GeV$ \ \ \ \ \ \ \ \ \ \ \ \ \
\ \ \ \ \ \ \ \ \ \ \ \ \ \ \ \ \ \ \ \ \ \ \ \ (28)

At about $t$ $\sim $ 10$^{-34}$ s following the beginning of the expansion,
it held

$\alpha _{g}=\alpha _{w}$ \ \ \ \ \ \ \ \ \ \ \ \ \ \ \ \ \ \ \ \ \ \ \ \ \
\ \ \ \ \ \ \ \ \ \ \ \ \ \ \ \ \ \ \ \ \ \ \ \ \ \ \ \ \ \ \ \ \ \ \ (29)

and treating (3), (13) and (29) we get in an independent way an expression
for $m_{x}$

$m_{x}=m_{Pc}.\alpha _{w}^{1/2}\cong 10^{16}GeV$ \ \ \ \ \ \ \ \ \ \ \ \ \ \
\ \ \ \ \ \ \ \ \ \ \ \ \ \ \ \ \ \ \ \ \ \ (29)

We supposed the values $m_{x}$ represent the mass of vector bosons $x$ and $y
$.

Conclusions

1. The present contribution rationalizes the ``constant'' $\alpha _{g}$ and
makes possible to prove on a exact base some relations describing our
Universe that have been up to now formulated as empirical postulates and
formulae.

2. The contribution corrects the Dirac presumption on a time decrease of the
constant $G$.

3. Using simple relations the contribution determines in an independent way
the mass of

\ \ vector bosons $x$ and $y$.

References

[1] I. L. Rozentahl, Adv. Math. Phys. Astr. 31 (1986) 241

[2] J. Sima, M. Sukenik, General Relativity and
Quantum Cosmology,

\ \ \ \ http://xxx.lanl.gov/abs/qr-qc/9903090 (1999)

[3] V. Skalsky, M. Sukenik, Astrophys. Space Sci., 178 (1991)
169

\end{document}